\documentstyle[aps,prl,twocolumn,psfig]{revtex}
\begin{document}
\frenchspacing
\title{Generalised-Lorentzian Thermodynamics}
\author{Rudolf A. Treumann}
\address{Max-Planck-Institute for extraterrestrial Physics, Postfach 1603, 
D-85748 
Garching, Germany 
and \\
International Space Science Institute, Hallerstrasse 6, CH-3012 Bern, 
Switzerland \\
(tre@mpe.mpg.de or treumann@issi.unibe.ch)}
\draft
\maketitle
\begin{abstract}
 
\end{abstract}
We extend the recently developed non-gaussian thermodynamic formalism \cite{tre98} of a (presumably strongly turbulent) non-Markovian medium to its most general form that allows for the formulation of a consistent thermodynamic theory. All thermodynamic functions, including the definition of the temperature, are shown to be meaningful. The thermodynamic potential from which all relevant physical information in equilibrium can be extracted, is defined consistently. The most important findings are the following two: (1) The temperature is defined exactly in the same way as in classical statistical mechanics as the derivative of the energy with respect to the entropy at constant volume. (2) Observables are defined in the same way as in Boltzmannian statistics as the linear averages of the new equilibrium distribution function. This lets us conclude that the new state is a real thermodynamic equilibrium in systems capable of strong turbulence with the new distribution function replacing the Boltzmann distribution in such systems. We discuss the ideal gas, find the equation of state, and derive the specific heat and adiabatic exponent for such a gas. We also derive the new Gibbsian distribution of states. Finally we discuss the physical reasons for the development of such states and the observable properties of the new distribution function. 

\section{Introduction}
\noindent
In a recent paper \cite{tre98} we developed the kinetic theory of a collisionless (presumably non-Markovian) equilibrium state of a system of $N$ particles undergoing strongly turbulent interactions. Our motivation was to investigate what kind of statistical mechanics described an equilibrium state of fully developed stationary turbulence, if it existed at all. Evidence of the possibility of such a description comes from observations of L\'evy flights and started in the twentieth of this century when Richardson found his turbulent spectral law that was fundamentally different from Kolmogorov's law of ordinary spectral behaviour in turbulence (for a review of the history see, e.g., \cite{shl93}). 

There had been earlier attempts to construct other than Maxwell-Boltzmann probability distributions already in the past century and continued until today, mostly as attempts to describe the thermodynamics of extreme situations \cite{lav95}. These attempts culminated in a purely mathematical extension of Boltzmann's definition of entropy by R\'enyi \cite{bal56,ren70}. These extensions, though widely used in chaotic dynamics in order to infer about multi-fractal behaviour, have not been given any physical justification yet. A certain mathematical simplification of R\'enyi's original proposal has subsequently been suggested \cite{tsa88} that seemed to have applicability to some range of physical problems. In particular, some kind of thermodynamics was constructed (for the most lucid presentation see, e.g., \cite{bog96}). However, this theory refers to an unusual prescription of constructing physical observables that is not in agreement with conventional physics. 

Our basic assumption was that the stationary turbulence we were going to describe was not describable by weak turbulence theory. This assumption implied that any equilibrium state the system might have achieved could not be constructed by the means of perturbation technique, i.e. no small expansion parameter exists. Systems of this kind are subject to non-perturbation techniques. In some of those systems the transition from the original state towards turbulence proceeds through criticality. Such systems are critical and must be treated by renormalisation group methods (cf., e.g., \cite{wil83,cha92}). We did not explicate about this point but instead asked for the properties of the corresponding equilibria, proposing that the system had actually settled into a critical and turbulent intermediate equilibrium. We found that these equilibrium states were describable by a new {\it non-Boltzmannian} collision integral
\begin{equation}
{\cal C}_{\mathrm T}=\int {\rm d}\tau\frac{{\rm d}\sigma_{\mathrm T}}{{\rm d}\Omega}\,{\rm d}\Omega \, G_{\mathrm T}[12],
\label{eq1}
\end{equation}
where $\sigma_{\mathrm T}$ is the (turbulent) collisional cross-section, and
\begin{equation}
G_{\mathrm T}[12]\equiv g[f(1^\prime)]g[f(2^\prime)]-g[f(1)]g[f(2)]
\label{eq2}
\end{equation}
is the {\it correlation functional} of the one-particle distribution functions $f$ after (primed) and before (unprimed) the interaction. $G_{\mathrm T}[12]$ itself was found to be the product of functionals $g[f]$ each depending on the distribution function of one family of particles only. (The index ${\mathrm T}$ indicates that the systems are in a state of about stationary turbulence.) The turbulent state in that they are found is reached on passing through critical points after having entered a highly nonlinear phase. The transition to the turbulence is not known in detail, but it can be assumed that during the critical phase the systems evolves into all scales such that the scales cannot be separated anymore. It becomes essentially scale-invariant, and any perturbation theory breaks down when all scales are highly correlated. In \cite{tre98} it had been demonstrated that the functionals $g[f]$ actually consist of infinite products of correlations suggesting that this interpretation is close to the truth. 

The breakdown of perturbation theory in the scale-invariant state suggests that a microscopic approach to the problem will have to refer to renormalization group techniques. We have shown in \cite{tre98}, however,  that it is possible to describe the stationary equilibrium state of the system without the need to develop the microscopic phase-transition theory in detail by the methods of statistical mechanics. This has been done in close analogy to Boltzmann's kinetic theoretical approach. Detailed balance then requires that
\begin{equation}
\ln g[f]=-\beta(\epsilon_{\bf p}-\mu)
\label{eq3}
\end{equation}
where $\epsilon_{\bf p}=p^2/2m$ is the energy of a particle of momentum ${\bf p}$ and mass $m$, and $\beta, \mu$ are two arbitrary constants (playing the role of Lagrangean multipliers) which have been suggested \cite{tre98} to correspond to the kinetic temperature and chemical potential of the system, respectively. Below this suggestion will be proved in full strength.

If the system is described by Equation (\ref{eq1}) and is in the assumed highly nonlinear multi-scale turbulent equilibrium, then one can introduce \cite{tre98} a control parameter $\kappa$ such that
\begin{equation}
\lim\limits_{\kappa\to\infty} g[f(\kappa,\epsilon_{\bf p})]\to f_{\mathrm B}(\epsilon_{\bf p})
\label{eq4}
\end{equation}
reproduces the Boltzmann distribution function $f_{\mathrm B}$, and $G_{\mathrm T}$ becomes the ordinary Boltzmann collision functional. In \cite{tre98} a particular functional $g[f]$ was found and it was demonstrated that this functional actually described a thermal equilibrium state nicely satisfying an $H$-theorem and permitting for a new expression for the turbulent entropy ${\cal S}_{\,\mathrm T}$. One should note that this new mathematical expression for the entropy does not attach any new physical interpretation to the entropy.  As in the Boltzmann case, entropy describes the amount of irreversible disorder in the system. The new expression for the entropy merely means that in the turbulent scale-invariant state of the system the increase of disorder is calculated in a different way than in the conventional Boltzmann state. 

\section{Generalisation}
\noindent
We now generalise the functional $g[f]$ to its most general form
\begin{equation}
g[f(\kappa, \epsilon_{\bf p})]=\exp\left\{ \kappa\left[1-f^{-1/(\kappa+\ell)}(\kappa,\epsilon_{\bf p})\right]\right\}
\label{eq5}
\end{equation}
where $\{ \kappa, \ell \} \in \textsf{\textbf  R}$, and $\ell$ is an arbitrary {\it fixed} real number ($\textsf{\textbf R}$ is the space of real numbers). The advantage of introducing $\ell$ will become clear below. It is then easy to demonstrate that the condition (\ref{eq4}) is satisfied for {\it any} arbitrary fixed $\ell\neq\infty$.

The equilibrium distribution function $f(\kappa, \epsilon_{\bf p})$ can be constructed using the two equations (\ref{eq3}) and (\ref{eq5}). The most general distribution function is found to be
\begin{equation}
f_\ell(\kappa, \epsilon_{\bf p})= \left(1-\frac{\beta\mu}{\kappa}+\frac{\beta\epsilon_{\bf p}}{\kappa}\right)^{-(\kappa+\ell)}.
\label{eq6}
\end{equation}
Clearly, it is a function of particle momentum ${\bf p}$ through the particle energy $\epsilon_{\bf p}=p^2/2m$ and of the two parameters, $\kappa$ and $\ell$, respectively. It is then easy to show in parallel to \cite{tre98} that $f_\ell$ satisfies the following new generalised turbulent entropy relation referred to in the above discussion:
\begin{equation}
{\cal S}_{\,{\mathrm T},\ell}=-k_{\mathrm B}V\int  \frac{{\mathrm d}^3p}{h^3}\,f_\ell(\epsilon_{\bf p},\kappa) \ln g[f_\ell(\epsilon_{\bf p},\kappa)].
\label{eq7}
\end{equation}
This entropy is concave and moreover is super-additive, meaning that the entropy of two independent systems is {\it larger} than the sum of the individual entropies of the two systems, an interesting property of such states.
(Note that, in accord with physical intuition and requirement, ${\cal S}_{\,{\mathrm T}, \ell}$ can never become sub-additive. Entropy in a single closed system or in a collection of closed systems will always grow as disorder cannot be extinguished by adding other disorder.) That this is true can be shown along the same lines as in \cite{tre98}, for the introduction of the fixed number $\ell$ does not introduce any change in the analysis. Hence, $f_\ell$ is an actual thermodynamic equilibrium distribution that replaces the Boltzmann distribution $f_{\mathrm B}$ under $\kappa$-conditions. All the discussion of \cite{tre98} can be applied to it.

In the following we develop the corresponding thermodynamics and show that it requires the choice $\ell=1$. Therefore, the correct {\it thermodynamic equilibrium} one-particle distribution function of the $\kappa$-gas is given by
\begin{equation}
f(\epsilon_{\bf p},\kappa)= \left(1-\frac{\beta\mu}{\kappa}+\frac{\beta\epsilon_{\bf p}}{\kappa} \right)^{-(\kappa+1)},
\label{eq8}
\end{equation}
where, for convenience, we suppressed the index $\ell=1$.

\section{Thermodynamic Relations}
\noindent
In this section we define the basic thermodynamic functions and demonstrate that a $\kappa$-gas is a system that though behaving in a special way is nevertheless in complete accord with the fundamental thermodynamic relations. 

\subsection{Thermodynamic Potentials}
\noindent
All macroscopic thermodynamic information about a system in contact with the external world is contained in the thermodynamic potential ${\cal Q} (\beta, V,\mu)$ which is a function of the temperature variable $\beta$, the volume of the system, $V$, and the chemical potential $\mu$ (or particle number $N$). We define ${\cal Q}_{\mathrm T}(\beta, V, \mu)$ of the $\kappa$-gas by
\begin{equation}
{\cal Q}_{\mathrm T}(\beta,V,\mu)=-\frac{V}{\beta}\int\frac{{\mathrm d}^3p}{h^3}\, \left(1-\frac{\beta\mu}{\kappa} +\frac{\beta\epsilon_{\bf p}}{\kappa}\right)^{-\kappa}.
\label{eq9}
\end{equation}
Since we assume that thermodynamics should provide a valid description of the macrostate of the $\kappa$-gas in order to be in accord with conventional physics, the relation between the thermodynamic potential and other thermodynamic functions is given by
\begin{equation}
{\cal Q}_{\mathrm T}=F_{\mathrm T}-\mu N.
\label{eq10}
\end{equation}
Here $F_{\mathrm T}$ is the Helmholtz free energy of the $\kappa$-gas, and ${\cal Q}_{\mathrm T}$ is the difference between $F$ and the product of the average particle number and the chemical potential. Hence the free energy is given by
\begin{equation}
F_{\mathrm T}={\cal Q}_{\mathrm T}+\mu N=-\frac{V}{\beta}\int\frac{{\mathrm d}^3p}{h^3}\left[1-\frac{\beta\mu}{\kappa} +\frac{\beta\epsilon_{\bf p}}{\kappa}\right]^{-\kappa}+\mu N.
\label{eq11}
\end{equation}
We now show that this definition is consistent with the Boltzmann limit. Indeed, taking the limit $\kappa\to\infty$ in the integral we immediately find that
\begin{equation}
F=-\frac{V}{\beta}\int\frac{{\mathrm d}^3p}{h^3}\exp\left[-\beta(\epsilon_{\bf p}-\mu)\right]+\mu N,
\label{eq12}
\end{equation}
which coincides with the Boltzmann-Helmholtz free energy. Moreover, this expression identifies $\mu$ with the chemical potential, and $\beta=1/k_{\mathrm B}T$ with the inverse kinetic temperature. The latter expression will be proved explicitly and in full generality for arbitrary $\kappa$ below.

\subsection{Average Energy and Number Density}
\noindent
In conventional statistical mechanics, the total mean energy $E$ is calculated from the Helmholtz free energy in the following way:
\begin{eqnarray}
E&=&\frac{\partial}{\partial\beta}(\beta F)\nonumber\\
&=&-\frac{\partial}{\partial\beta}\left[V\int\frac{{\mathrm d}^3p}{h^3}\exp\left[-\beta(\epsilon_{\bf p}-\mu)\right]-\mu N\right].
\label{eq13}
\end{eqnarray}
In the new thermodynamics we replace $F$ with $ F_{\,\mathrm T}$ to obtain
\begin{eqnarray}
E&=&\frac{\partial}{\partial\beta}(\beta F_{\,\mathrm T})\nonumber\\
&=&-\frac{\partial}{\partial\beta}\left[V\int\frac{{\mathrm d}^3p}{h^3}\left(1-\frac{\beta\mu}{\kappa} +\frac{\beta\epsilon_{\bf p}}{\kappa}\right)^{-\kappa}-\mu N\right].
\label{eq13a}
\end{eqnarray}
Carrying out the partial differentiation,
this yields the following expression for the energy 
\begin{eqnarray}
E&=&V\int\frac{{\mathrm d}^3p}{h^3}\frac{\epsilon_{\rm p}}{(1-\beta\mu/\kappa +\beta\epsilon_{\bf p}/\kappa)^{\kappa+1}}\nonumber\\
&=&V\int\frac{{\mathrm d}^3p}{h^3}\epsilon_{\bf p}f(\epsilon_{\bf p},\kappa).
\label{eq14}
\end{eqnarray}
Note that the second part of this equation is just the correct physical definition of the average energy of the system as the integral over phase space of the distribution function, Equation (\ref{eq8}), as is required by the commonly used definition of an observable in statistical mechanics and kinetic theory. In this respect, our theory is thus consistent with common statistical physics and kinetic theory. It does not require any different kind of averaging in order to calculate the observables as the physically relevant quantities. This implies that this theory is also in accord with the fundamental kinetic BBGKY theory (cf., e.g., \cite{lib79}, chp. 2). 

In order to be consistent with the previous sections, Equation (\ref{eq14} suggest that we must identify $\ell=1$, which justifies our previous choice for $\ell$. It would of course be possible to choose any arbitrary $\ell$ in the definition of ${\cal Q}$ and to adjust the exponent of $f$ appropriately, but such an action would introduce some unnecessary arbitrariness that would result in a simple re-scaling of $\kappa$. 

The average particle number $N$ is then given as the zeroth order moment of the distribution function $f(\epsilon_{\bf p},\kappa)$. This can be shown to be the negative of the partial derivative of the above thermodynamic potential ${\cal Q}_{\mathrm T}$
\begin{eqnarray}
N&=&V\int\frac{{\rm d}^3p}{h^3}\left[1-\frac{\beta\mu}{\kappa}+\frac{\beta\epsilon_{\bf p}}{\kappa} \right]^{-(\kappa+1)}\nonumber\\
&=&-\left(\frac{\partial{\cal Q}_{\mathrm T}}{\partial\mu}\right)_{\beta V}.
\label{eq15}
\end{eqnarray}
The average density of the $\kappa$-gas is correspondingly obtained as
\begin{equation}
n\equiv\frac{N}{V}=
=\int \frac{{\mathrm d}^3p}{h^3}\,f(\epsilon_{\bf p},\kappa).
\label{eq16}
\end{equation}
On the other hand, it must be required that the chemical potential of the $\kappa$-gas is the partial derivative of the free energy $F_{\mathrm T}$ with respect to the average particle number
\begin{equation}
\mu=(\partial F_{\mathrm T}/\partial N)_{\beta V}.
\label{eq17}
\end{equation}
This equation is conventionally used to express $\mu$ through $N$. 

\subsection{Entropy}
\noindent
In order to find the entropy relation, we form the following derivative of the thermodynamic potential $k_{\mathrm B}(\partial{\cal Q}_{\mathrm T}/\partial\beta^{-1})_{V\mu}$. It can be shown that this derivative can be written as
\begin{eqnarray}
k_{\mathrm B}\left[\frac{\partial {\cal Q}_{\mathrm T}}{\partial(1/\beta)}\right]_{V\mu}
&=& -k_{\mathrm B}\beta [E-\mu N -{\cal Q}_{\mathrm T}]\nonumber\\
&=&\frac{1}{T}(F_{\mathrm T}-E).
\label{eq18}
\end{eqnarray}
The last expression is just the negative of the entropy ${\cal S}_{\,\mathrm T}$ and coincides with the following definition
\begin{eqnarray}
{\cal S}_{\,\mathrm T}&=&-k_{\mathrm B}V\int\frac{{\mathrm d}^3p}{h^3} \,f(\epsilon_{\bf p}, \kappa)\ln g[f(\epsilon_{\bf p}, \kappa)]\nonumber\\ 
&=&-k_{\mathrm B}V
\int\frac{{\mathrm d}^3p}{h^3}\,f(\epsilon_{\bf p},\kappa)\left[1-f^{-1/(\kappa + 1)}(\epsilon_{\bf p},\kappa)\right]
\label{eq19}
\end{eqnarray}
when replacing the distribution function inside the integral with Equation (\ref{eq8}) and using the above derived representations for the average energy (\ref{eq14}), particle number (\ref{eq15}), and the thermodynamic potential (\ref{eq9}). 

Clearly this entropy expression has a structure different from the ordinary Boltzmann definition. It contains the logarithm of the correlation functional $g[f]$ in place of the logarithm of the distribution function itself. This fact implies that in the $\kappa$-state of the gas it is the correlations that contribute most to the entropy. While in the final state of the system when the interactions become purely stochastic and the gas settles into its thermal death, $g\to f_{\mathrm B}$, and the entropy assumes its classical representation.

\subsection{Consequence I: Definition of Temperature}
\noindent
As an important application of the above definition of the entropy ${\cal S}_{\,\mathrm T}$ of the $\kappa$-gas we now derive the expression for the thermodynamic temperature. From classical thermodynamics it is known that the only consistent definition of the temperature is given in the form of a derivative of the entropy:
\begin{equation}
\frac{1}{T}=\left(\frac{\partial{\cal S}}{\partial E}\right)_{V\mu}.
\label{eq20}
\end{equation}
When we use the expression (\ref{eq8}) for the distribution function in the definition of the entropy (\ref{eq19}) we recover that
\begin{equation}
{\cal S}_{\,\mathrm T} =k_{\mathrm B}\beta \,(E-\mu N),
\label{eq21}
\end{equation}
which is nothing else thatn a rearranged version of (\ref{eq18}).
Then taking the partial derivative with respect to the average energy $E$ we immediately identify the temperature $T$ as in conventional thermodynamics with the inverse of the Lagrangean multiplier $\beta$
\begin{equation}
\frac{1}{\beta}=k_{\mathrm B}T.
\label{eq22}
\end{equation}

This very important relation proves that the temperature $T$ of the $\kappa$-gas is defined exactly in the same manner as the temperature of the classical Boltzmann gas. In this way it renders all other definitions of the temperature of the $\kappa$-gas used in the literature invalid. Those definitions still contained dependencies on $\kappa$ (cf., e.g., \cite{mey89,tho91,mac95,mac98} and elsewhere). These $\kappa$-dependencies turn out to be unphysical. They are the result of a naiv use of the so-called experimentally determined `$\kappa$-distributions' in calculating a formal expression for the temperature.  The physically correct distribution function Equation (\ref{eq8}) resembles the $\kappa$-distribution, but it contains the non-vanishing chemical potential. It is only this complete distribution function that leads to a correct thermodynamics, with the temperature $T$ being defined as a physical quantity by its thermodynamic definition, Equations (\ref{eq20}) and (\ref{eq22}). The temperature in this sense is a measure of the state of the system. It is a parameter that characterises the gas. It is not a measure of the mean energy. Only in the Boltzmann gas the two, mean energy and temperature, measured in energy units, are related in a simple way. Below we are going to derive the relation valid in the $\kappa$-gas.  

The above derivation of the temperature thus justifies the use of the constant $\beta$ as the unambiguous and only measure of the thermodynamic temperature of any (turbulent) $\kappa$-system and shows that in such a system the above definition of the entropy consistently replaces the Boltzmann definition.

\subsection{Consequence II: Equation of State}
\noindent
In order to complete the set of thermodynamic relations we may now construct the equation of state taking the partial derivative of ${\cal Q}_{\mathrm T}$ with respect to the volume $V$. This procedure yields the pressure
\begin{eqnarray}
P &=&-\left(\frac{\partial{\cal Q}_{\mathrm T}}{\partial V}\right)_{\beta\mu}\nonumber\\
&=&\frac{1}{\beta}\int \frac{{\mathrm d}^3p}{h^3}\left(1-\frac{\beta\mu}{\kappa}+\frac{\beta\epsilon_{\bf p}}{\kappa}\right)^{-\kappa}.
\label{eq23}
\end{eqnarray}
Rewriting this expression with the help of Equation (\ref{eq9}) we find the following important relation
\begin{equation}
P V=-{\cal Q}_{\mathrm T}(\beta, V, \mu).
\label{eq24}
\end{equation}
This is the fundamental equation of state of a (turbulent) $\kappa$-gas. It shows that such gases possess complicated equations of state. Such behaviour has been expected from the very beginning, because the presence of the long-range correlations should become manifest in the average properties of the gas as well. Even an ideal $\kappa$-gas turns out to have a non-simple equation of state, as will be demonstrated below. 

The remaining first and second order thermodynamic relations can all be obtained from the previous relations and will not be given here.

\section{Review of Ideal $\kappa$-Gas Properties}
\noindent
In our previous paper \cite{tre98} we derived some of the relations for an ideal gas. Here, because of the precise definition of the thermodynamic potential and the above consistency proof of the thermodynamic relations we can considerably simplify the expressions given before. Moreover, the formulas given below correct them for thermodynamic consistency.

\subsection{Chemical Potential}
\noindent
We start with the average particle number $N$ of an ideal $\kappa$-gas. This number is given as the phase-space integral over the ideal gas distribution function $f(\epsilon_{\bf p},\kappa)$ Equation (\ref{eq8})
\begin{eqnarray}
N&=&V\frac{\Gamma(\kappa -1/2)}{\Gamma(\kappa+1)}\nonumber\\
&\times&\left(\frac{2\pi m\kappa}{h^2\beta}\right)^{3/2}\left(1-\frac{\beta\mu}{\kappa}\right)^{-(\kappa-1/2)}.
\label{eq25}
\end{eqnarray}
It is assumed throughout thermodynamics that the average number density $n=N/V$ would be known. Hence, Equation (\ref{eq25}) is basically an equation for the chemical potential $\mu$ of the ideal $\kappa$-gas. Inverting (\ref{eq25}) and introducing the `quantum $\kappa$-density' 
\begin{equation}
n_{q\kappa}=\left(\frac{2\pi m\kappa}{h^2\beta}\right)^{3/2}
\label{eq26}
\end{equation}
we obtain for the chemical potential
\begin{equation}
1-\frac{\beta\mu}{\kappa}=\left[\frac{n_{q\kappa}}{n}\frac{\Gamma(\kappa-1/2)}{\Gamma(\kappa+1)}\right]^{1/(\kappa-1/2)}.
\label{eq27}
\end{equation}
The right-hand side of this expression is always positive, and hence either the chemical potential is negative $\mu<0$ or $\beta\mu/\kappa<1$. It is not too difficult to demonstrate \cite{tre98} that in the limit $\kappa\to\infty$ this expression reproduces the classical chemical potential of an ideal gas \cite{hua87} with $n_q=(2\pi m/h^2\beta)^{3/2}$ the `quantum density' of the classical gas. 

\subsection{Mean Energy}
\noindent
Also, carrying out the integration in Equation (\ref{eq14}), the average energy of the ideal $\kappa$-gas follows as
\begin{eqnarray}
E &=& \frac{3N}{2\beta} \frac{\kappa}{\kappa-3/2}\left(1-\frac{\beta\mu}{\kappa}\right) \nonumber \\
&=& \frac{3N}{2\beta}\frac{\kappa}{\kappa-3/2} \left[\frac{n_{q\kappa}}{n}\frac{\Gamma(\kappa-1/2)}{\Gamma(\kappa+1)}\right]^{1/(\kappa-1/2)}.
\label{eq28}
\end{eqnarray}
With the above identification of $\beta=1/k_{\mathrm B}T$ one immediately realises that the limit $\kappa\to\infty$ reproduces the classical well-known result (cf., e.g., \cite{hua87}, p.\ 77)
\begin{equation}
\frac{E}{N}=\frac{3}{2}k_{\mathrm B}T
\label{eq29}
\end{equation}
for the energy per particle. As is obvious from Equation (\ref{eq28}), in the $\kappa$-gas this last relation between the mean energy and the temperature becomes much more involved. Moreover, it is also obvious that the above definition of the energy requires that
\begin{equation}
\kappa > \kappa_{\mathrm min}=\frac{3}{2},
\label{eq29a}
\end{equation}
a condition that is consistent with a similar one derived in our previous publication \cite{tre98}.

\subsection{Ideal Gas Equation of State}
We now calculate the pressure of the ideal $\kappa$-gas in order to determine its equation of state. From Equation (\ref{eq24}) we obtain
\begin{eqnarray}
PV&=&\frac{N}{\beta}\frac{\kappa}{\kappa-3/2}\left(1-\frac{\beta\mu}{\kappa}\right)\nonumber \\
&=&\frac{N}{\beta}\frac{\kappa}{\kappa-3/2}\left[\frac{n_{q\kappa}}{n}\frac{\Gamma(\kappa-1/2)}{\Gamma(\kappa+1)}\right]^{1/(\kappa-1/2)}.
\label{eq30}
\end{eqnarray}
This equation of state becomes the ordinary ideal gas equation of state $PV=Nk_{\mathrm B}T$ only in the Boltzmann limit $\kappa\to\infty$. The ideal $\kappa$-gas behaves differently, possessing a much more complicate equation of state. This seems reasonable because, as mentioned above, the fact that the $\kappa$-gas is a highly correlated many-body system, which we assume is in an evolved turbulent state, must in the first place become obvious in its equation of state. Such a gas should exhibit a behaviour that differens from that of an ordinary laminar ideal gas. Nevertheless, however, is is then both surprising and satisfactory that, comparing Equations (\ref{eq28}) and (\ref{eq30}), one recovers the basic thermodynamic relation between the pressure and average energy, 
\begin{equation}
PV=\frac{2}{3}E,
\label{eq31}
\end{equation}
as is valid also in ordinary ideal gases. The unbroken validity of this relation even in the $\kappa$-gas is interesting and is an important finding. Its physical meaning is that it relates the mean energy in a simple geometrical way to the pressure of the $\kappa$-gas. This retains the physical interpretation that the pressure is the geometrical effect of the average motion of the constituents of the gas. The temperature assumes quite a different meaning then as an independent parameter measuring the irreversibility of the system in place of its thermal mean energy. The average energy of the $\kappa$-gas is thus not a simple measure of the temperature $T$ of the $\kappa$-gas. The average energy per particle $E/N$ in a $\kappa$-gas is not anymore a simple fraction of the kinetic temperature. Only in the Boltzmann gas become both, mean energy and temperature, simple equivalents. 

\subsection{Specific Heat and Adiabatic Index}
\noindent
In order to complete the discussion of the ideal gas, we calculate the specific heat $C_V$ at constant volume $V$ from
\begin{equation}
C_V=-\beta^2k_{\mathrm B}\,\left(\frac{\partial E}{\partial\beta}\right)_{V\mu}.
\label{eq32}
\end{equation}
Since $E$ is known, it is simple matter to find
\begin{eqnarray}
C_V&=&3k_{\mathrm B}N\,\left(\frac{\kappa}{\kappa-3/2}\right)\nonumber\\
&\times&\left\{ \left[\frac{n_{q\kappa}}{n} \frac{\Gamma(\kappa-1/2)}{\Gamma(\kappa+1)}\right]^{1/(\kappa-1/2)}-\frac{1}{2}\right\}.
\label{eq33}
\end{eqnarray}
Again, in the limit $\kappa\to\infty$, this expression converges to the ideal gas value $(3/2)k_{\mathrm B}N$ as the quantity in the curly brackets becomes just 
\begin{equation}
\lim_{\kappa\to\infty}(2\pi m/h^2\beta)^{3/(2\kappa -1)}\to 1. 
\label{eq34}
\end{equation}
However, as expected, $C_V$ in the $\kappa$-gas is not a constant (or otherwise an extensive function of the particle number). It depends on both, density and temperature. This complication suggests that the adiabatic exponent $\gamma_\kappa$ of the $\kappa$-gas will as well not be a simple constant, as is the case for the ideal Boltzmann gas.

Calculation of the specific heat $C_P$ at constant pressure turns out to be more involved. We can, however, take advantage of the general thermodynamic formula
\begin{equation}
C_P-C_V=T\,\frac{(\partial P/\partial\beta)_V}{(\partial\beta/\partial V)_P},
\label{eq35}
\end{equation}
relating $C_P$ and $C_V$ (cf., e.g., \cite{hua87}). This is possible because we have already demonstrated that the thermodynamics of the $\kappa$-gas is formally identical with that of the ordinary Boltzmann gas. 

Making extensive use of the equations of state (\ref{eq30}), energy (\ref{eq28}), and specific heat at constant volume (\ref{eq33}) we obtain for the difference (\ref{eq35}) of the two specific heats 
\begin{eqnarray}
C_P-C_V&=&k_{\mathrm B}N\frac{\kappa}{\kappa-3/2}\nonumber\\ &\times&\left[\frac{n}{n_{q\kappa}}\frac{\Gamma(\kappa +1)}{\Gamma(\kappa -1/2)}\right]^{1/(\kappa-1/2)}.
\label{eq36}
\end{eqnarray}
Inspection of this equation suggests that in a dilute $\kappa$-gas where the density $n=N/V\ll n_{q\kappa}$ is much less than the quantum density, the difference between the two specific heats will become small. On the other hand, close to maximum correlation (at $\kappa\to 3/2$) this difference can become large. These are two further distinctions between ordinary and $\kappa$-gases, respectively. 

The most interesting quantity is, however, the index of adiabaticity, $\gamma=C_P/C_V$. Dividing by $C_V$ in (\ref{eq36}) one finds that in the $\kappa$-gas this ratio becomes
\begin{eqnarray}
\gamma_\kappa&=&1+\frac{1}{3}\left[ \frac{n}{n_{q\kappa}}\frac{\Gamma(\kappa +1)}{\Gamma(\kappa-1/2)}\right]^{1/(\kappa-1/2)} \nonumber\\
&\times&\left\{ \left[\frac{n_{q\kappa}}{n} \frac{\Gamma(\kappa-1/2)}{\Gamma(\kappa+1)}\right]^{1/(\kappa-1/2)}-\frac{1}{2}\right\}^{-1}.
\label{eq37}
\end{eqnarray}
Indeed, for $\kappa\to\infty$ one can show that 
\begin{equation}
\lim\limits_{\kappa\to\infty}\, \gamma_\kappa=\frac{5}{3},
\label{eq38}
\end{equation}
which is in accord with the thermodynamics of an ordinary (three-dimensional) gas. But, for $\kappa <\infty$, the adiabatic index may deviate considerably from this value. In particular, when the term in the brackets containing the quantum density is sufficiently large compared to $1/2$, 
\begin{equation}
\gamma_\kappa\approx 1 + \frac{1}{3}\left[\frac{n}{n_{q\kappa}}\frac{\Gamma(\kappa + 1)}{\Gamma(\kappa-1/2)}\right]^{2/(\kappa-1/2)}.
\label{eq39}
\end{equation}
This value is clearly smaller than $5/3$ and in the limit of a very dilute gas approaches unity. In particular, taking $\kappa=\kappa_{\mathrm min}=3/2$, 
\begin{equation}
\gamma_{3/2}\approx 1+\frac{\pi}{18}\left(\frac{n}{n_q}\right)^2.
\label{eq39a}
\end{equation}
This is very close to unity for small ratios $n/n_q$. 
For classical $\kappa$-gases we can therefore conclude that the adiabatic index falls into the intervall
\begin{equation}
1 < \gamma_\kappa \leq \gamma = \frac{5}{3}
\label{eq40}
\end{equation}
with the right-hand side holding for the ordinary gas. On the other hand, there is the possibility that in very dense $\kappa$-gases with $n\sim n_{q\kappa}$ the second term in Equation (\ref{eq37}) becomes negative. In such a situation, values $\gamma_\kappa < 1$ would even arise. For instance, assuming $n\to n_q$, one finds that
\begin{equation}
\lim\limits_{n\to n_q}\gamma_{3/2}=1-\frac{\sqrt{2}\pi}{3\sqrt{3}-2\sqrt{2}}<0
\label{eq40a}
\end{equation}
becomes negative. Of course, this extreme case is physically impossible, and is a result of using the minimum value for $\kappa$, and $n=n_q$ at the same time. This discussion shows, however, that for sufficiently high densities and small $\kappa$ the adiabatic index becomes small.
Investigation of these cases requires the development of the quantum theoretical extension of the present theory. This will be given elsewhere \cite{tre99}.

The above result on the reduction of the adiabatic index is exciting. Reduction of $\gamma$ physically implies that the number of degrees of freedom increases. Remember that, classically, $\gamma$ is a measure of the internal freedom of the system and is microscopically related to the degrees $d$ of freedom by 
\begin{equation}
\gamma=(d+2)/d.
\label{eq41}
\end{equation}
Equation (\ref{eq40}) then suggests that for $\kappa_{\mathrm min}<\kappa<\infty$, the number $d$ increases by a large amount. For $d\gg 2$, the index $\gamma$ quickly approaches the value $\gamma\to 1$. In the $\kappa$-gas, $\gamma_\kappa$ may not necessarily be related to $d$ in the same simple way (\ref{eq41}) as in the classical gas. It is possible that the relation between $\gamma_\kappa$ and $d$ will be more involved. In order to obtain the actual dependence, one must develop the microscopic theory of the interactions in a multi-scale medium which is outside our present reaches. Nevertheless, the decrease in $\gamma$ provides a strong argument for the validity of the initial assumption underlying our theory. This assumption was that the $\kappa$-gas is in a multi-scale state with an enormously large number of degrees of freedom, and that these scales are all correlated in a way as is expected for fully developed turbulence. Media like this are believed to be scale-invariant and should microscopically be treated by renormalisation-group methods \cite{cha92}.

\subsection{Isentropic Process}
\noindent
With the adiabatic index at hand we can investigate the isentropic (adiabatic) evolution of the $\kappa$-gas. To this end we need the explicit expression for the ideal gas entropy
\begin{eqnarray}
{\cal S}_{\,{\mathrm T}}&=&k_{\mathrm B}V\left( \frac{2\pi\kappa m}{h^2\beta}\right)^{3/2}\frac{\Gamma (\kappa-3/2)}{\Gamma(\kappa)}\nonumber\\
&\times&\left\{ \frac{1}{2}+\left[\frac{n_{q\kappa}}{n}\frac{\Gamma(\kappa- 1/2)}{\Gamma(\kappa+1)}\right]^{1/(\kappa-1/2)}\right\}.
\label{eq42}
\end{eqnarray}
Isentropic processes leave the entropy constant. To first approximation this requires that
\begin{equation}
VT^{3/2}={\mathrm const},
\label{eq43}
\end{equation}
a condition that is identical to the condition for adiabaticity in the thermodynamics of ordinary gases (cf., e.g., \cite{kit80}). For $\kappa$-gases, the second term in the curly brackets in Equation (\ref{eq42}) seems to introduce a further complication. However, in the particular case when the term in the square brackets on the right-hand side of Equation (\ref{eq42}) is large, we still arrive at the surprising result that the condition Equation (\ref{eq43}) is still exactly satisfied. Hence, only the intermediate regime when
\begin{equation}
\frac{n_{q\kappa}}{n}\sim \frac{\Gamma(\kappa+1)}{2^{\kappa-1/2}\Gamma(\kappa-1/2)}
\label{eq43a}
\end{equation}
the adiabatic relation between volume and temperature changes. This practical independence of the condition of isentropy on the value of $kappa$ is a purely geometrical effect that tells that a sufficiently fast expansion of any medium should cause cooling. It is only reasonable that this behaviour does not halt in front of turbulent or scale-invariant systems. With the help of (\ref{eq43}) the two remaining isentropic relations are obtained from the equation of state as
\begin{eqnarray}
PT^{-5/2}&=&{\mathrm const},\\
PV^{5/3}&=&{\mathrm const}.
\label{eq43b}
\end{eqnarray}
It is remarkable that these conditions exactly coincide with those for the isentropic processes in a classical Boltzmann gas. We thus learn from this agreement that isentropy results in simple geometric behaviour of both Boltzmann and $\kappa$-gases. Since the thermodynamic differential equations remain valid for the $\kappa$-gas, they can in principle be used to express these exponents through $\gamma_\kappa$. In a $\kappa$-gas, however, the ratio of specific heat has no direct relation to the isentropic processes as is the case in an ordinary ideal gas.  

\subsection{Volume Coefficients}
\noindent
In order to conclude this section we finally derive the coefficients of thermal expansion and compressibility, respectively. The first is given by the well-know formula
\begin{equation}
\alpha_{\mathrm ex}=\frac{1}{V}\left(\frac{\partial V}{\partial T}\right)_{P\mu}.
\label{eq44}
\end{equation}
It can be easily calculated from the equation of state (\ref{eq30}) of the $\kappa$-gas. Replacing the chemical potential in the final expression after having performed the relevant differentiations, we find that
\begin{equation}
\alpha_{{\mathrm ex},\kappa}=\frac{1}{T}\left[ \frac{n}{n_{q\kappa}} \frac{\Gamma(\kappa +1)}{\Gamma(\kappa-1/2)}\right]^{1/(\kappa-1/2)}
\label{eq45}
\end{equation}
retains its inverse proportionality to the temperature $T$ while otherwise being small for dilute media.
The thermal compressibility 
\begin{equation}
{{\mathrm K}}_T=-\frac{1}{V}\left(\frac{\partial V}{\partial P}\right)_{T\mu}
\label{eq46}
\end{equation}
can most easily be obtained from Equation (\ref{eq36}) exploiting the relation
\begin{equation}
C_P-C_V= TV \frac{ \alpha_{\mathrm exp}}{{\mathrm K}_T}.
\label{eq47}
\end{equation}
This relation together with Equation (\ref{eq45}) leads to the following useful result
\begin{eqnarray}
{\mathrm K}_{T\kappa}&=&\frac{V}{Nk_{\mathrm B}T}\frac{\kappa-3/2}{\kappa}\nonumber\\ &\times&\left[\frac{n}{n_{q\kappa}}\frac{\Gamma(\kappa+1)}{\Gamma(\kappa-1/2)}\right]^{1/(\kappa -1/2)}.
\label{eq48}
\end{eqnarray}
This completes our account of the stationary properties of the ideal $\kappa$-gas. 

\section{Fluctuations}
\noindent
Because it is a system in thermal equilibrium the $\kappa$-gas is capable of thermal fluctuations as well. Such fluctuations are known to be the root-mean-square amplitudes of the oscillations of the various macroscopic quantities around the thermal equilibrium state. Independent of the very particular form of the equilibrium distribution function the fluctuation amplitudes average out when averaged over sufficiently long times or spatial scales. Under $\kappa$-conditions this restriction poses caution on the definition of fluctuations. Times can be no longer than the inverse binary collision time as it is our philosophy that the collisionless turbulent scale-invariant state refers to times shorter than the binary collision time $1/\nu_c$ (cf., \cite{tre98}). Hence, very slow oscillations may not average out even linearly and may survive the entire scale-invariant regime until they enter the binary collisional state when they become ultimately depleted. Moreover, the assumption of scale-invariance (or self-similarity) also implies that very long oscillations may survive. With these restrictions in mind the rms fluctuation amplitude of a quantity $A$ is defined in the ordinary way as
\begin{equation}
\langle \Delta A\rangle_{\mathrm rms}^2=\langle A^2\rangle-\langle A\rangle^2
\label{eq50}
\end{equation}
where the average is understood as the ensemble average or expectation value. Our discussion presented in the previous sections has shown that calculating expectation values by linear averaging over the distribution function is the appropriate way to stay in accord with the requirement of being able to define a consistent thermodynamic theory in the $\kappa$-regime of the evolution of the system. Here we are interested in providing the expressions for the most fundamental rms fluctuation amplitudes.

\subsection{Particle Number Fluctuations}
\noindent
The particle number fluctuation cannot be calculated directly from the zeroth order moment of the distribution function. Remembering, however, that the particle number is closely related to the chemical potential the general thermodynamic expression for the number fluctuation is (cf., e.g., \cite{hua87} p.\ 152) 
\begin{equation}
\langle\Delta N\rangle_{\mathrm rms}^2=k_{\mathrm B}TV\,\frac{\partial^2 P}{\partial\mu^2}.
\label{eq51}
\end{equation}
The second-order derivative of the pressure on the right-hand side of this equation can be shown \cite{hua87} to be equivalent to
\begin{equation}
\frac{\partial^2 P}{\partial\mu^2}=-\frac{N^3}{V^3}\left[\frac{\partial P}{\partial (V/N)}\right]^{-1}.
\label{eq52}
\end{equation}
Using the ideal gas equation (\ref{eq30}) for the pressure, we obtain for the rms particle number fluctuation
\begin{eqnarray}
\frac{\langle\Delta N\rangle_{\mathrm rms}^2}{N^2}&=&
\frac{\kappa-3/2}{N\kappa}\nonumber\\
&\times&\left[\frac{n}{n_{q\kappa}}\frac{\Gamma(\kappa+1)}{\Gamma(\kappa-1/2)}\right]^{1/(\kappa-1/2)}.
\label{eq53}
\end{eqnarray}
The rms number fluctuation increases with $N$ while the relative fluctuation in particle number decreases as $N$ increases.
Note that if we would have used the average expressions to calculate the number fluctuation, we would have found a zero fluctuation in number. Hence the effect is small in a dilute gas, as it should be. Moreover, comparing the fluctuation with the compressibility (\ref{eq48}) one observes that the number fluctuation is proportional to ${\mathrm K}_{T\kappa}$ as has been expected since this is a particular case of the fluctuation-dissipation theorem holding for small fluctuations (cf., e.g., \cite{deg84}, chpt. 8).

\subsection{Energy Fluctuations}
It is more interesting to determine the fluctuation of the mean energy. It is most convenient to simply calculate the two expectation values in Equation (\ref{eq50}) with $A\equiv E$ and to take their difference. It is, however, much simpler to use the general thermodynamic expression for the energy fluctuation
\begin{equation}
\langle\Delta E\rangle_{\mathrm rms}^2=-(\partial E/\partial\beta)
\label{eq54}
\end{equation}
in order to obtain the average fluctuation of the internal mean energy
\begin{equation}
\langle\Delta E\rangle_{\mathrm rms}^2=\frac{3}{2}\left(\frac{\kappa}{\kappa-3/2}\right)\,N(k_{\mathrm B}T)^2.
\label{eq55}
\end{equation}
The rms fluctuation amplitude is proportional to the thermal energy $k_{\mathrm B}T$ and increases as the root of the particle number. Its absolute value also increases with $\kappa$ approaching $\kappa_{\mathrm min}$. 
The smaller $\kappa$ the higher is the absolute level of the fluctuations in energy. On the other hand, when dividing the rms amplitude by the mean energy Equation (\ref{eq28}) one finds that at the same time the relative energy fluctuation 
\begin{equation}
\frac{\langle\Delta E\rangle_{\mathrm rms}^2}{E^2}\propto \frac{1}{N}\frac{\kappa-3/2}{\kappa}
\label{eq56}
\end{equation}
decreases because the mean energy increases faster than the fluctuation. This can be easily understood as the increasing effect of storage of energy in the high-energy tail of the distribution near $\kappa_{\mathrm min}$. The relative amount of energy stored in the fluctuations produced by the tail becomes less than the energy itself. Nevertheless, though the energy diverges faster than the fluctuation energy, the gas contains fluctuations of large amplitude that is an expression of its turbulent nature and the appearance of many scales.
 
\section{Properties of the Equilibrium Distribution}
\noindent
In this section we return to the equilibrium distribution function $f(\epsilon_{\bf p}, \kappa)$ and investigate its most obvious properties. Because distribution functions nowadays have become measurable quantities (though not observables in the strict physical sense) their shape and behaviour provide a direct link to the physical properties of the medium. Examples of so-called $\kappa$-distributions observed in space plasmas have been given continuously for about thirty years (cf., \cite{vas68,scu79,cri88,cri91,lin95}), with the most extended observations being obtained in the Earth's magnetospheric tail \cite{cri88,cri91}. Similar distributions are believed to be related to L\'evy flight dynamics \cite{shl93} and have been used in an attempt to interpret cross-field diffusion \cite{tre97}. But the physics of such $\kappa$-distributions has not been understood for a long time. Attempts of their explanation trace back to various types of solutions of the Fokker-Planck equation for particular interactions \cite{has85,col95,rey98}. The general thermodynamic theory developed in \cite{tre98} and in the present paper is the first consistent physical theory of these families of distribution functions, aside from another approximate attempt \cite{tsa95} that was based on so-called 
non-extensive thermodynamics \cite{tsa88}.

In the following we discuss the information that can be extracted from observation of generalised-Lorentzian distribution functions.

\subsection{Velocity Distribution}
\noindent
The distribution function $f(\epsilon_{\bf p},\kappa)$ Equation \ref{eq8} has been given in terms of the particle momentum ${\bf p}$. It belongs to the family of generalized-Lorentzian functions. Distribution functions like this one possess the property that not all infinitely many moments of the distribution can exist. For given $\kappa$, the moments of order $r>2\kappa-1$ start diverging. This fact restricts the value of $\kappa$ to those values that satisfy the condition
\begin{equation}
\kappa>(r+1)/2.
\label{eq57}
\end{equation}
For the mean energy to be a physically measurable quantity this expression sets the limit for $\kappa>3/2$. At the current state it is difficult to discuss what will happen to the higher moments and in which way the theory can be `renormalised' if at all. Probably, the high energy tails will break the scale-invariance at some stage and will close the system of moments by self-limitation of the tail of the distribution function. The more important requirement here is that $f(\epsilon_{\bf p}, \kappa$) is a probability distribution function and must therefore be positive and real. This property renders
\begin{equation}
\beta\mu/\kappa <1,\qquad 0\leq\epsilon_{\bf p}<\infty
\label{eq58}
\end{equation}
a condition that has been shown to be satisfied for ideal gases and in the absence of external potential fields. Otherwise one requires that
\begin{equation}
\epsilon_{\bf p}>{\tilde\mu}-\kappa/\beta>0,
\label{eq59}
\end{equation}
where ${\tilde\mu}$ is the total chemical potential including the external potential field. In the second case, a certain range of low particle energies (or low particle velocities) is not covered by the distribution. This implies that for sufficiently large positive total chemical potentials the moment integrals may become non-analytic. Such cases provide enormous difficulties for physical interpretation. In the following we will exclude this case from
discussion.

In terms of the velocity ${\bf v}$ of the particles the isotropic velocity distribution function reads
\begin{equation}
f({\bf v},\kappa)=\left(1-\frac{\beta\mu}{\kappa}+\frac{m\beta v^2}{2\kappa}\right)^{-(\kappa+1)}.
\label{eq60}
\end{equation}
Since $1-\beta\mu/\kappa>0$, there exists a range of velocities for which
\begin{equation}
m\beta v_c^2/2\kappa <1-\beta\mu/\kappa.
\label{eq61}
\end{equation}
In this range of velocities the velocity distribution function
\begin{equation}
f({\bf v},\kappa)\sim (1-\beta\mu/\kappa)^{-(\kappa+1)}={\mathrm const}, \qquad v<v_c
\label{eq62}
\end{equation}
is practically {\it flat}, a property that has been frequently observed but was barely understood so far. The present theory provides a simple straightforward interpretation of this flatness problem. 

On the other hand, for high velocities $v\gg v_c$
the distribution function becomes power law
\begin{equation}
f({\bf v},\kappa)\propto v^{-2(\kappa+1)}.
\label{eq63}
\end{equation}
Any fit at high velocities with exponent $\alpha$ will provide a measurement of $\kappa=\alpha/2 -1$. The break point of the distribution is at velocity
\begin{equation}
v_b\approx [2\kappa(1-\beta\mu/\kappa)/m\beta]^{1/2},
\label{eq64}
\end{equation}
allowing for the determination of $\beta=1/k_{\mathrm B}T$, i.e. the determination of the temperature. 

A general important property of the (isotropic) velocity distribution is that the value $4\pi p^2 f({\bf p},\kappa)$ is the probability of finding any particles in the intervall of momentum $[p,p+{\mathrm d}p]$. The maximum of this expression thus defined the most probable velocity ${\bar v}$ (momentum) of the particles. This velocity is given by
\begin{equation}
{\bar v} = \left(\frac{2k_{\mathrm B}T}{m}\right)^{1/2}\left[\frac{n_{q\kappa}}{n} \frac{\Gamma(\kappa-1/2)}{\Gamma(\kappa+1)}\right]^{1/(2\kappa-1)}.
\label{eq65}
\end{equation}
On the other hand, the rms speed is defined as 
\begin{equation}
v_{\mathrm rms}=\frac{1}{N}\left[\frac{4\pi V}{h^3}\int\,{\mathrm d}^3p\, v^2 f({\bf v},\kappa)\right]^{1/2}.
\label{eq66}
\end{equation}
The value obtained is
\begin{equation}
v_{\mathrm rms}= = \left(\frac{3k_{\mathrm B}T}{m}\right)^{1/2} \left(\frac{\kappa}{\kappa-3/2}\right)\left[\frac{n_{q\kappa}}{n} \frac{\Gamma(\kappa-1/2)}{\Gamma(\kappa+1)}\right]^{1/(2\kappa-1)}.
\label{eq67}
\end{equation}
It is not surprising that this value differs from ${\bar v}$ as this is also true for the Maxwell-Boltzmann distribution of velocities. The ratio of both speeds is
\begin{equation}
\frac{{\bar v}}{v_{\mathrm rms}}=\sqrt{\frac{2}{3} }\left(\frac{\kappa-3/2}{\kappa}\right).
\label{eq68}
\end{equation} 
The most probable speed is smaller than the rms velocity. The value of these expressions is mainly that measurement of both velocities provides an independent possibility to determine the value of $\kappa$ and, subsequently, the value of the thermodynamic temperature $T$.

\subsection{Energy Distribution}
\noindent
We briefly turn to a discussion of the energy distribution as this in practice is sometimes more important than the velocity distribution itself. Similar to the Boltzmann case the energy distribution is defined as
\begin{equation}
f({\epsilon,\kappa})=\epsilon^{1/2}f({\bf v},\kappa).
\label{eq69}
\end{equation}
At small particle energies $\epsilon$ it increases as $f\propto\epsilon^{1/2}$,
Reaches maximum at
\begin{equation}
\frac{\beta{\bar\epsilon}}{\kappa}=\frac{1-\beta\mu/\kappa}{2\kappa+1},
\label{eq70}
\end{equation}
and, at high energies, becomes power-law and drops as $\epsilon^{\kappa+1/2}$. Again, from its asymptotic behaviour one determines $\kappa$, while the value of the most probable energy is a measure of $\beta\mu$. One also notes that the most probable energy {\it increases} with temperature $T$. The maximum value of the distribution itself is
\begin{equation}
f_{\mathrm m}({\bar\epsilon},\kappa)=\frac{(2\kappa m^3)^{1/2}}{[2(\kappa+1)]^{\kappa+1}}\left(\frac{2\kappa+1}{1- \beta\mu/\kappa}\right)^{\kappa+1/2}.
\label{eq71}
\end{equation}
This value is always smaller than $\sqrt{m^3/\beta}$. Hence this value increases at about $\sqrt{T}$. Its determination allows for the measurement of the thermodynamic temperature of the gas and in combination with the above formula also for the measurement of the chemical potential.

\subsection{Gases in External Potentials}
\noindent
An important frequently realised case is that a gas is brought into an external potential field $\phi$. Then the distribution becomes
\begin{equation}
f(\epsilon_{\bf p}-\phi,\kappa)=\left[1-\frac{\beta\mu}{\kappa}+\frac{\beta(\epsilon_{\bf p}-\phi)}{\kappa}\right]^{-(\kappa+1)}.
\label{eq72}
\end{equation}
The external potential simply changes the value of the chemical potential
\begin{equation}
\mu\to\mu+\phi
\label{eq73}
\end{equation}
as in an ordinary gas as well. Depending on the sign of the external potential this may have enormous consequences on the analyticity of the distribution function, as has been mentioned in the introduction to this section. As long as this analyticity is guarantied, further integration of the distribution function provides no difficulties.

As for a first and simple application we calculate the density of an electron plasma immersed into an external electric field. In this case the average external density in the absence of the field is $n_0$, and $\phi\to -e\phi$ in Equation (\ref{eq72}) is understood as the potential energy of the electron in the electric potential field $\phi$. Carrying out the integration of the zeroth order moment of \ref{eq72} one finds for the electron density
\begin{equation}
\frac{n(\phi)}{n_0}=1-\frac{e\phi}{k_{\mathrm B}T}\frac{\kappa-1/2}{\kappa} \left[\frac{n_0}{n_{q\kappa}}\frac{\Gamma(\kappa+1)}{\Gamma(\kappa-1/2)} \right]^{1/(\kappa-1/2)}.
\label{eq74}
\end{equation}
This expression replaces the well-known commonly used Boltzmann formula
\begin{equation}
n_{\mathrm B}(\phi)=n_0\exp(e\phi/k_{\mathrm B}T).
\label{eq75}
\end{equation}
Clearly, this expression is the limiting form of Equation (\ref{eq74}) for $\kappa\to\infty$.
An exactly equivalent formula may be found, e.g., for gases embedded in an external gravitational field of acceleration $-{\bf g}$ (the so-called barometric law). In order to obtain the corresponding expression one replaces $e\phi/k_{\mathrm B}T\to -mgH/k_{\mathrm B}T$, where $H$ is the height above level $H=0$, where $n(H=0)=n_0$. The interesting point about both these density distributions is that, in a $\kappa$-gas, the density reacts algebraically to the presence of the external potential. For the barometric case this implies that it decays less than exponentially.

A simple application of the density formula Equation (\ref{eq74}) is to the calculation of the screening distance of an ion in a quasineutral $\kappa$-plasma, the so-called Debye length. This length may be found, expanding the right-hand side of Equation (\ref{eq74}) for small potentials and using Poisson's law. For illustration we show only the one-dimensional case. Then, from
\begin{equation}
\frac{{\mathrm d}^2\phi}{{\mathrm d}x^2}=-\frac{e\,\Delta n}{\in_0}
\label{eq76}
\end{equation}
(with $\in_0$ the vacuum dielectric constant) we find that the new Debye length $\lambda_{{\mathrm D},\kappa}$ can be expressed through the Boltzmann-Debye length $\lambda_{\mathrm DB}$ as
\begin{equation}
\lambda_{{\mathrm D},\kappa}^2= \lambda_{\mathrm DB}^2\frac{\kappa}{\kappa-3/2} \left[\frac{n_{q\kappa}}{n_0}\frac{\Gamma(\kappa-1/2)}{\Gamma(\kappa+1)} \right]^{1/(\kappa-1/2)}.
\label{eq77}
\end{equation}
This value always exceeds the Boltzmann-Debye length. Intuitively this effect is clear and very satisfactory is as far as one may easily convince oneself that the presence of an excess of electrons in the tail of the distribution function over the Maxwell-Boltzmann distribution will worsen the conditions for screening. The faster electrons are less affected by the potential of the test ion and, hence, this potential will reach farther out into space. Actually, at the absolute limiting value $\kappa_{\mathrm min}=3/2$ the ratio of the two Debye-lengths becomes practically infinite,
\begin{equation}
\lim\limits_{\kappa\to 3/2}\left(\frac{\lambda_{{\mathrm D},\kappa}}{ \lambda_{\mathrm DB}}\right)\to\left(\frac{4^{1/3}3\pi k_{\mathrm B}T}{n_0h^2} \right)^{3/4}.
\label{eq78}
\end{equation}
The effect of such large screening lengths on the behaviour of $\kappa$-plasmas is of considerable interest. It may possibly support electric fields in plasmas stronger than ordinarily expected to exist. Also, because the level of electron thermal fluctuations in Langmuir waves is proportional to $1/n_0\lambda_{{\mathrm D},\kappa}^3$, Langmuir fluctuations in $\kappa$-plasmas should be strongly reduced. It may be speculated that such plasmas would be even less collisional and exhibit lower levels of spontaneous emission as well as lower levels of thermal radiation from the plasma. However, no firm conclusion can be drawn at this level of investigation as long as the microscopic processes acting in $\kappa$-plasmas have not been clarified.

\subsection{Relativistic Distribution}
\noindent
In this last subsection we briefly investigate the relativistic version of the Loorentzian distribution function. In relativistic gases, $\epsilon_{\bf p}=m\gamma_{\mathrm rel}c^2$, with $\gamma_{\mathrm rel}$ the relativistic Gamma-factor. The phase space volume element transforms as
\begin{equation}
4\pi{\mathrm d}^3x\,p^2{\mathrm d}p=4\pi\, V_0m^3c^3(\gamma_{\mathrm rel}^2-1)^{1/2}{\mathrm d}\gamma_{\mathrm rel},
\label{eq79}
\end{equation}
where ${\mathrm d}V={\mathrm d}V_0/\gamma_{\mathrm rel}, p=m\gamma_{\mathrm rel}=mc(\gamma_{\mathrm rel}^2-1)^{1/2}$. This leads to the following expression for the thermodynamic potential
\begin{equation}
{\cal Q}=-\frac{4\pi\,V_0m^3c^3}{h^3\beta}\int\limits_1^\infty\,
\frac{{\mathrm d}\gamma_{\mathrm rel} (\gamma_{\mathrm rel}^2-1)^{1/2}
}{[1-\beta\mu/\kappa+(\beta mc^2/ \kappa )\gamma_{\mathrm rel}]^{\kappa} }.
\label{eq80}
\end{equation}
With the help of this expression we can calculate the particle number
\begin{eqnarray}
N&=&-\left(\frac{\partial{\cal Q}}{\partial\mu}\right)_{\beta V_0}\nonumber\\
&=&\frac{4\pi\,V_0m^3c^3}{h^3}\int\limits_1^\infty\,\frac{{\mathrm d}\gamma_{\mathrm rel}(\gamma_{\mathrm rel}^2-1)^{1/2}}{[1-\beta\mu/\kappa+(\beta mc^2/\kappa)\gamma_{\mathrm rel}]^{\kappa+1}}.
\label{eq80a}
\end{eqnarray}
From this formula we identify the relativistic distribution function as
\begin{eqnarray}
f_{\mathrm rel}(\gamma_{\mathrm rel},\kappa)&=& \frac{4\pi\,V_0m^3c^3}{h^3}(\gamma_{\mathrm rel}^2-1)^{1/2}\nonumber\\
&\times&\left[1-\frac{\beta\mu}{\kappa}+\frac{\beta mc^2}{\kappa}\gamma_{\mathrm rel}\right]^{-(\kappa+1)}, \qquad \kappa>2.
\label{eq80b}
\end{eqnarray}
The restriction on $\kappa$ results from the requirement that the relativistic energy as a moment of the distribution function must exist. For small but still relativistic energies, i.e., for 
\begin{equation}
1<\gamma_{\mathrm rel}< (\kappa/\beta mc^2)(1-\beta\mu/\kappa)
\label{eq81}
\end{equation}
the relativistic distribution function varies according to
\begin{equation}
f_{\mathrm rel}(\gamma_{\mathrm rel},\kappa)\sim\sqrt{2}(\gamma_{\mathrm rel}-1)^{1/2}.
\label{eq82}
\end{equation}
This increase with energy reproduces the $\sqrt{\epsilon_{\bf p}}$ {\it low energy} increase of the non-relativistic energy distribution function found above. On the other hand, in the ultra-relativistic domain $\gamma_{\mathrm rel}\gg 1$,
\begin{equation}
f_{\mathrm rel}(\gamma_{\mathrm rel},\kappa)\propto \gamma_{\mathrm rel}^{-\kappa},
\label{eq83}
\end{equation}
implying a flatter decay than exhibited by the non-relativistic energy distribution, the latter evolving as $f(\epsilon,\kappa)\propto \epsilon^{-(\kappa+1/2)}$. The maximum of this distribution is at energy
\begin{equation}
{\bar\gamma}_{\mathrm rel}= \frac{1-\beta\mu/\kappa}{2\beta mc^2}\nonumber\\
\left[ 1+\left(1+\frac{\kappa+1}{\kappa} \frac{4\beta^2m^2c^4}{1-\beta\mu/\kappa } \right)^{1/2}\right].
\label{eq84}
\end{equation}
This is the most probable relativistic energy in a $\kappa$-gas. 

\section{Discrete Energy Levels: Gibbsian Distribution}
\noindent
The theory developed in \cite{tre98} and explicated in the former sections of this paper assumes that the energy levels in the $\kappa$-gas are continuously distributed. If, on the other hand, the energy levels are discrete, the formalism must be replaced by a discrete formalism. This can be achieved by replacing the integrals in the most fundamental formulas by sums over the discrete states of the system. The most important replacement is to be done in the thermodynamic potential ${\cal Q}$ of Equation (\ref{eq9}). When summing over discrete states, the volume factor $V$ disappears, and the restriction on momentum space volume elements contained in the differential ${\mathrm d}^3p/h^3$ becomes unnecessary. Thus, the discrete {\it partition function} becomes
\begin{equation}
{\cal Q}_{\mathrm T}=-\frac{1}{\beta}\sum\limits_{r=1}^N\left(1-\frac{\beta\mu}{\kappa}+\frac{\beta E_r}{\kappa}\right)^{-\kappa}.
\label{eq85}
\end{equation}
Here $E_r$ is the $r$-th discrete energy level, and the summation is over all dynamically possible levels $E_r$. In a $\kappa$-system, this expression replaces (up to an classically unimportant factor containing the chemical potential) the classical (Gibbsian) partition function
\begin{equation}
{\cal Q}=\sum\limits_{r=1}^N\,\exp(-\beta E_r),
\label{eq86}
\end{equation}
which is the sum over Gibbs factors, as can be shown by forming the limit value for $\kappa\to\infty$. (Note that the above mentioned factor $-\beta^{-1}\exp(-\beta\mu)$ that remains afterwards drops out by simple normalisation in the classical case, while it is important to be retained in the $\kappa$-system.) The partition function (\ref{eq85}) contains all the information about the equilibrium state of the $\kappa$-system. In analogy to our derivation of the distribution function it also suggests that the most probable Gibbs distribution of discrete states can be written as
\begin{equation}
w_r=\left(1-\frac{\beta\mu}{\kappa}+\frac{\beta E_r}{\kappa}\right)^{-(\kappa+1)}.
\label{eq87}
\end{equation}
Here $w_r$ is a probability (or else the most probable occupation number of state $r$). Accordingly, the entropy of the discrete system becomes
\begin{equation}
{\cal S}_{\,\mathrm T}= -k_{\mathrm B}\kappa\sum\limits_{r=1}^N\, w_r(1-w_r^{-1/(\kappa+1)}),
\label{eq88}
\end{equation}
an expression that is of similar though not identical form as the one given in \cite{tsa88}. All thermodynamic relations apply to ${\cal Q}_{\mathrm T}$. It is important to note that there is no simple probabilistic way of distributing $N$ particles on arbitrary sates that would reproduce the above Lorentz-Gibbsian probability distribution. The occupation number $w_r$ is not a simple most probable probability distribution in the ordinary meaning of the word. It cannot be achieved at by throwing the dice. Though it is obvious that it maximises ${\cal S}_{\,\mathrm T}$ thereby leading to the state of maximum disorder, this does not mean that this state is achieved in a simple probabilistic way. There are non-obvious underlying processes that speed up this distribution. These processes will have to be investigated in future. 

\section{Conclusions}
\noindent
In the present paper we extended our initial theory of the statistical mechanics of the $\kappa$-state to develop the thermodynamics of such a state. We found that it is indeed possible and in asymptotic agreement with classical statistical mechanics to construct the full thermodynamics of the $\kappa$-state. It was possible to define the thermodynamic potential ${\cal Q}_{\mathrm T}$ that contains all the physics of the equilibrium state. One of the most important findings was that the definition of temperature can be retained in its full validity also in the $\kappa$-state of a system. Another not less important conclusion was that the definition of observables in the $\kappa$-state is exactly the same as in classical statistical mechanics. This fact is most satisfactory because it does not violate physical intuition. Moreover, the entropy can only grow in states of the kind corresponding to $\kappa$-states. This is satisfactory as well: addition of disorder does not diminish disorder, it can only increase it. There is no process in closed systems, independent on their internal dynamics that could minimise entropy. To this end one must do external work on the system. The claim that instability would generate order in a closed system and reduce entropy is misleading. Entropy can only be reduced locally on the expense of entropy increase in other locations in the system. Instability can do nothing else but redistribute free energy until it makes it available to other dissipative processes. 

A large number of questions has been left open for future investigation. The most important is not that concerning the reality of processes like the once discussed here. Observation of L\'evy flights and increasing frequency of measurement of so-called `$\kappa$-distributions' (after those we have modelled our distribution function by the use of the control parameter $\kappa$) in large collisionless systems like the near-Earth space environment provide sufficient justification for an expedition into basic physics. But the question remains of what is the physical relevance of the index $\kappa$? In an interesting paper \cite{has85}, Hasegawa and co-workers found a similar distribution function solving a Fokker-Planck equation under the assumption of a very particular weakly turbulent interaction in a plasma. There, $\kappa$ turned out to depend on the dispersion of plasma waves. 

A solution of this kind may remain a particular case of weak turbulence. It rather seems that the $\kappa$ state of a system refers to an intermediate strongly turbulent state when the system has settled into quasi-equilibrium for some time before binary collisions set on to destroy the state and to dissipate the energy stored in the turbulent motion, as has been argued in \cite{tre98}.    
Equilibria usually imply the dissipative destruction of the free energy. In $\kappa$-equilibrium it seems that this is not necessarily the case. But the observation that $\kappa$ appears only in this intermediate state, suggests that there must be discontinuous transitions between the initial and the $\kappa$-state and possibly also to the Boltzmann regime. 

\begin{figure}[ht]
\centerline{\psfig{figure=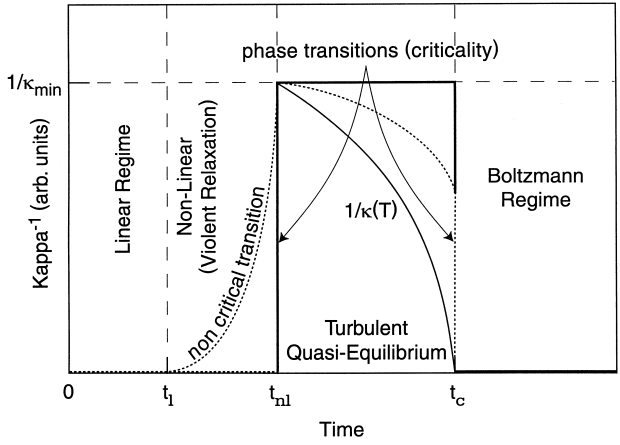,width=8.4cm,clip=}}
\caption{Schematic evolution of $1/\kappa$ for a different cases. 
The solid line refers to the model used in the present paper. $\kappa$ is zero
At times $t<t_{nl}$ and $t>t_c=1/nu_c$, respectively. There are two critical transitions in this case where $\kappa$ jumps from zero to a finite value $\kappa>\kappa_{\mathrm min}$ at these transitions (the figure shows the marginal case $\kappa=\kappa_{\mathrm min}$). The dotted line shows a possible
Non critical transition where $\kappa\neq 0$ already in the interval $t_l<t<t_{nl}$. Such a transition may avoid criticality and would be describable by ordinary strong turbulence theory (nonlinear wave-wave and wave-particle interactions). In addition, $\kappa$ can depend on time (or otherwise temperature if accounting for transitional heating in turbulence. Then either the cases of the dotted (retaining a late critical phase transition) or thin solid lines (no final critical transition) apply, respectively. 
\label{kappa}}
\end{figure}

Figure \ref{kappa} schematically illustrates this behaviour of $\kappa$ for the same scenario as in \cite{tre98}. In this figure we plotted the value of $\kappa^{-1}$. This value must be zero in the respective linear and in the final Boltzmann regimes, because these regimes are classical ($\kappa\to\infty$). In the turbulent quasi-equilibrium state $\kappa^{-1}$ suddenly assumes a value between zero and $1/\kappa_{\mathrm min}$ (the figure shows the marginal case $\kappa = {\mathrm const} = \kappa_{\mathrm min} $). The sudden increase from $\kappa^{-1}=0$ to $\kappa^{-1}\neq 0$ is understood as a critical phase transition. A possible non-critical transition is shown as the dotted line in the non-linear violent relaxation regime where our theory does not apply.

Physically spoken, there is no need for $\kappa$ being constant troughout the entire quasi-equilibrium. If $\kappa$ changes sufficiently slowly compared to the length of this quasi-equilibrium the equilibrium theory can still be applied. There is reason to believe that such a situation is closer to reality than $\kappa={\mathrm const}$. Dissipative non-collisional irreversible processes will necessarily accompany the turbulent state. Because the heating causes $T$ to increase with time, $\kappa(T)$ will itself become a function of time. Two such cases have been included in Figure \ref{kappa}. In the first one (thin solid line), the final critical phase transition does entirely disappear. $\kappa(T)$ tends to infinity right at the end of the intermediate stationary state. In the second case (dotted line), $\kappa$ remains finite towards the end of the quasi-equilibrium, and another critical phase transition is needed when suddenly binary collisions take over. Which one of these cases will be realised, depends on the properties of the system during its evolution. We will have to develop the non-equilibrium thermodynamics for $\kappa$-systems before these questions can be answered. This goal in addition requires to solve the problem of divergencies appearing in the higher moments of the distribution function. 

\section*{Acknowledgments}
It is a great pleasure to thank J. Geiss, B. Hultqvist (both ISSI) and G. Morfill (MPE) for their continuous interest and moral support during the period of writing this report. Discussions with T. Chang, J. Geiss, M. Hoshino, A. Kull, G. Morfill, M. Scholer, J. Scudder, and T. Terasawa are deeply acknowledged. Most of this work has been performed at the ISSI Bern, after it had originally been initiated during a visiting professorship 
at STEL, Nagoya University, Japan. The hospitality of Y. Kamide and S. Kokubun at STEL, Toyokawa, as well as of Nagoya University is also gratefully acknowledged.

\end{document}